\documentclass[
superscriptaddress,
preprint,
nofootinbib,
amsmath,amssymb,
aps,
prd,
]{revtex4-1}

\usepackage{graphicx}
\usepackage{amsmath}
\usepackage{amssymb}
\usepackage{bm}
\usepackage[usenames,dvipsnames]{color}
\usepackage{dcolumn}
\usepackage{hyphenat}
\usepackage{hyperref}
\usepackage{url}
\usepackage{enumitem}
\usepackage{subfigure}
\usepackage{soul}

\begin{document}

\title{A model-independent reconstruction of the matter power spectrum}

\author{Gen Ye}
\email{ye@lorentz.leidenuniv.nl}
\affiliation{Institute Lorentz, Leiden University, PO Box 9506, Leiden 2300 RA, The Netherlands}

\author{Jun-Qian Jiang}
\email{jiangjq2000@gmail.com}
\affiliation{School of Physical Sciences, University of Chinese Academy of Sciences, Beijing 100049, China}

\author{Alessandra Silvestri}
\email{silvestri@lorentz.leidenuniv.nl}
\affiliation{Institute Lorentz, Leiden University, PO Box 9506, Leiden 2300 RA, The Netherlands}

\begin{abstract}
We propose a new model-independent reconstruction method for the matter power spectrum based on its time dependence and a combination of observations from different redshifts. The method builds on a perturbative expansion in terms of the linear growth function, with each coefficient in the expansion being a free function of scale, to be reconstructed from the data. When using the linear growth function of a specific cosmological model, e.g. $\Lambda$CDM, the  reconstruction can serve as a consistency check for non-linear modeling in that given model, as well as a new method for detecting departures from the assumed model  in the data.
As an application, we show how using DES Y3 3x2pt and Planck PR4 CMB lensing data, assuming a $\Lambda$CDM linear growth and first order expansion, the reconstructed matter power spectrum $P_{\rm m}(k)$ is compatible with that computed from $\Lambda$CDM and halo model. In particular, we show that the method reconstructs the non-linear part of $P_{\rm m}(k)$ for $k\gtrsim 1\ \rm{Mpc}^{-1}$ without the need of assuming a non-linear model. 
\end{abstract}

\maketitle


\section{Introduction} \label{sec:intro}

The cosmological constant cold dark matter ($\Lambda$CDM) model has been very successful in fitting the state-of-art observations of background and linearly perturbed cosmology, especially the spectrum of temperature anisotropies in the cosmic microwave background (CMB)~\cite{Planck:2019nip,ACT:2020gnv,SPT-3G:2022hvq}, baryon acoustic oscillations (BAO)~\cite{eBOSS:2020yzd} and distance measurements from type Ia supernovae (SNIa)~\cite{Brout:2022vxf}. However, with the advent of precision cosmology, some inconsistencies have started to arise between different recent observations, when interpreted within $\Lambda$CDM. The most prominent one is the so-called Hubble tension, in which the Hubble constant $H_0$, characterizing the current expansion rate of the Universe, derived from CMB+BAO assuming $\Lambda$CDM differs from that measured locally by Cepheid calibrated SNIa at $5\sigma$~\cite{Riess:2019qba,Rubin:2023ovl}. Furthermore, the value of $S_8\equiv\sigma_8\sqrt{\Omega_m/0.3}$, $\sigma_8$ being the mean amplitude of linear matter perturbation in a sphere of radius $8h^{-1} \ \rm{Mpc}$, as measured by weak lensing surveys is $2-3\sigma$ smaller than that derived from  CMB+BAO within $\Lambda$CDM~\cite{KiDS:2020suj, DES:2021wwk, Kilo-DegreeSurvey:2023gfr}. More recently, the first BAO data release from DESI, combined with CMB and some recent type Ia data \cite{DES:2017myr,Rubin:2023ovl,Brout:2022vxf}, also shows a preference for dynamical dark energy over a cosmological constant \cite{DESI:2024mwx}, even hinting at non-minimally coupled gravity~\cite{Ye:2024ywg}. These inconsistencies, see e.g.~\cite{Efstathiou:2024dvn,Perivolaropoulos:2021jda,Abdalla:2022yfr} for some recent reviews, have triggered extensive discussions about testing $\Lambda$CDM against the newest data as well as explorations of  new physics that could explain these tensions. In this context, it is crucially important to develop model-independent approaches to the observations, on one hand to check the consistency with $\Lambda$CDM, on the other hand to thoroughly assess potential inconsistencies and link them to new physics. In this paper we develop such a method for the matter power spectrum. 

The linear dynamics of large scale structure (LSS) is well-understood in $\Lambda$CDM as well as in many  beyond $\Lambda$CDM models. With the help of the publicly available Einstein-Boltzmann solvers \texttt{CAMB}~\cite{Lewis:1999bs} and \texttt{CLASS}~\cite{Lesgourgues:2011re}, as well as their extensions such as~\texttt{EFTCAMB}~\cite{Hu:2013twa,Raveri:2014cka} and \texttt{hi\_class}~\cite{Zumalacarregui:2016pph,Bellini:2019syt} it is now possible to produce subpercent level theoretical predictions for all cosmological observables within the linear regime for $\Lambda$CDM and most dark energy and modified gravity models of interest (see e.g.~\cite{Bellini:2017avd}). However, interpreting observational data from galaxy  clustering and weak lensing~\cite{KiDS:2020suj,DES:2021wwk,eBOSS:2020yzd}, often requires modeling of the non-linear regime as well. This will be crucial for Stage IV LSS surveys that are deploying in these years~\cite{Laureijs2011,LSSTDarkEnergyScience:2018jkl}. Lacking a proper modeling of the non-linear regime, one has to resort to conservative scale cuts which result in throwing away a significant part of the data, (see  e.g.~\cite{DES:2017myr}). For cold dark matter, non-linear evolution can be studied via N-body simulations or emulators and fast approximate methods such as COLA, Bacco and Pinocchio,  see~\cite{Winther:2015wla, Winther:2019mus} for  overviews and comparisons of N-body codes and emulators available for $\Lambda$CDM and beyond. As we progress towards smaller scales, it becomes important to model also the baryonic feedback, either via some  prescription for the baryonic physics or with hydro-dynamical simulations 
; this is already complicated and computationally expensive for $\Lambda$CDM~\cite{vanDaalen:2011xb,Chisari:2019tus}, with the difficulties and uncertainties becoming larger for beyond $\Lambda$CDM models. Recently, it has been pointed out that the $S_8$ tension might originate from incomplete knowledge of the non-linear sector~\cite{Amon:2022azi}. Correspondingly, constructions of the full matter power spectrum agnostic to the exact non-linear physics have been proposed as an alternative way to interpret the data. For example, in~\cite{Amon:2022azi} it has been suggested to marginalizing over the non-linear correction with one parameter $A_{\rm mod}$ by setting $P_{\rm NL}=P_{\rm L}+A_{\rm mod}(P^{\rm theory}_{\rm NL}-P_{\rm L})$; this has later been generalized to a reconstruction of $P_{\rm NL}$ by promoting $A_{\rm mod}$ to a function of scale and redshift~\cite{Preston:2023uup,Preston:2024ggf}. In~\cite{Broxterman:2024oay}, the authors considered a reconstruction by expanding $P_m(k,a)$ in powers of $k$ and $a$.

We propose a new model-agnostic reconstruction of $P_{\rm{m}}(k,z)$ that exploits the different time dependence in its linear and non-linear parts, namely we write
\begin{equation}\label{eq:pk_expansion}
    P^{(n)}_{\rm{m}}(k,z) = \sum_{i=0}^n\alpha_i(k)D_g^{2(i+1)}(z_0,z) 
\end{equation}
where $\alpha_i$'s are $k$-dependent coefficients with the same dimension as $P(k)$ and $D_g(z_0, z)$ is the linear growth factor from redshift $z_0$ to $z$, with $z_0$ being some reference redshift. From a perturbation theory point of view, Eq.\eqref{eq:pk_expansion} represents the time dependence of perturbative loop expansion and one might compute the analytic form of $\{\alpha_i\}$ to finite order within a given cosmological model and non-linear prescription. For example, the effective field theory of large scale structure (EFTofLSS) \cite{Baumann:2010tm,Carrasco:2012cv} provides a systematic parameterization of $\{\alpha_i\}$. To this end, Eq.\eqref{eq:pk_expansion}, with ${\alpha_i}$ as free functions to reconstruct, is a more general series expansion of $P_{\rm m}$ in both $k$ and $z$ independent of non-linear modeling, and thus can serve as a consistency test for the non-linear prescriptions. In this paper we present a proof of concept study of this idea, showing that it can already reconstruct the linear and non-linear matter spectrum simultaneously at $k\gtrsim1 \ \rm{Mpc}^{-1}$ with current data. 

\section{Data and methodology}\label{sec:data}

\begin{figure}
\centering
\includegraphics{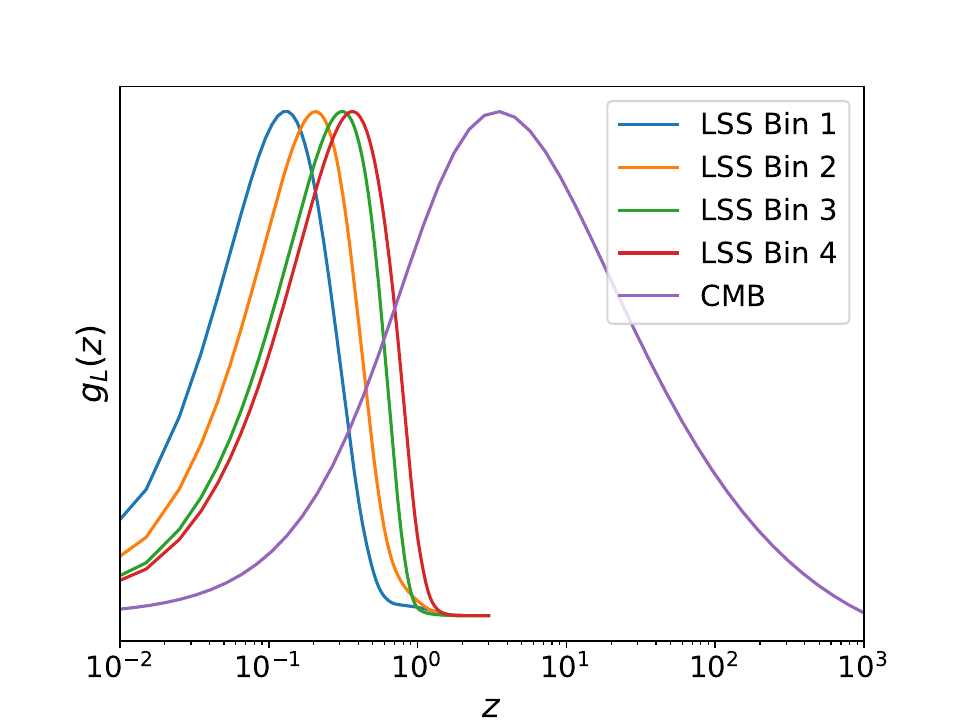}
\caption{Lensing kernel \eqref{eq:lensing kernel}, peak height normalized to unity, of the LSS and CMBL data. The four LSS bins correspond to the four tomographic bins of DES Y3 \cite{DES:2021bvc,DES:2021vln}.}
\label{fig:lensing kernel}
\end{figure}

We assume that the background cosmology is described by $\Lambda$CDM and use it to compute the linear growth rate $D_g(z_0, z)$ that goes into Eq.\eqref{eq:pk_expansion}. Physically, the leading order term $\alpha_0(k)D_g^2(z)$ corresponds to the linear matter power spectrum $P_{\rm{L}}(k,z)$. Therefore, we can re-parameterize $\alpha_0(k)\equiv A_0(k)P_{\rm{L}}(k,z_0)$ so that $A_0(k)$ is a dimensionless function centered around unity.  The zeroth order term in  Eq.\eqref{eq:pk_expansion} then becomes $A_0(k)P_{\rm{L}}(k,z)$.
Similarly one can define $\alpha_i(k)\equiv A_i(k)\left(\frac{P_{\rm{L}}(k,z_0)}{\rm{Mpc}^{-3}}\right)^i P_{\rm{L}}(k,z_0)$, which casts Eq.\eqref{eq:pk_expansion} into
\begin{equation}\label{eq:pk_expansion2}
    P^{(n)}_{\rm{m}}(k,z)=\sum_{i=0}^n A_i(k)\left(\frac{P_{\rm{L}}(k,z)}{\rm{Mpc}^{-3}}\right)^iP_{\rm{L}}(k,z)\,.
\end{equation}
Contrarily to $A_0$, the $A_i(k)$ with $i >0$ are not necessarily of order unity.
Eq.~(\ref{eq:pk_expansion2}) is the actual formula that will be used in the reconstruction. 

The starting point is the choice of a cosmological model, as the paradigm within which to calculate the linear power spectrum. One then can proceed with fitting~(\ref{eq:pk_expansion2}) to the dataset, modeling each $A_i(k)$ as a cubic interpolation over six nodes of function values $\ln A_i$ located at $\log_{10}(k_j \ \rm{Mpc})=[-3, -1.5, -1, -0.5, 0, 0.5]$, corresponding to $k_j =[0.001, \sim 0.03, 0.1, \sim 0.3, 1,\sim 3]\,{\rm Mpc}^{-1}$. The first node at $k_1=10^{-3} \ \rm{Mpc}^{-1}$, deeply in the linear regime, is used to anchor the corresponding $A_i(k)$ to its asymptotic value at $k\to0$.  Rigorously, one would need to optimize over the number of reconstruction nodes, e.g. using Bayes model comparison. For this initial exploration of the method, we simply tried 5-7 nodes and established that six nodes was a good balance between reconstruction flexibility and parameter number.
We compute the linear power spectrum $P_{\rm L}(k,z)$ using the cosmological code \texttt{CAMB}~\cite{Lewis:1999bs}. The full matter power spectrum $P_{\rm m}(k,z)$ is then reconstructed according to Eq.\eqref{eq:pk_expansion2} from the interpolated $\{A_i(k)\}$ given the nodes $\{\ln A_i(k_j)\}$. As we will see in Section-\ref{sec:result}, expanding to $n=1$ is sufficient for the data under consideration, since the latter can hardly constrain $n\ge2$. 

To reconstruct $P_{\rm m}(k,z)$ via Eq.\eqref{eq:pk_expansion2}, one needs observations of the matter power spectrum at different redshifts. To this end, we consider the following joint dataset: 
\begin{itemize}
    \item \textbf{LSS}: The 3x2pt galaxy shear and clustering observation from the data release 3 of the DES collaboration \cite{DES:2021wwk}.
    \item \textbf{CMBL}: The marginalized version of Planck PR4 CMB lensing reconstruction \cite{Carron:2022eyg} which only requires the CMB lensing spectrum as input.
    \item \textbf{SN}: Light curve measurements of 1550 type Ia supernova compiled in the Pantheon+ data \cite{Brout:2022vxf}.
    \item \textbf{BAO+RSD}: BAO measurement from MGS \cite{Beutler:2011hx}, 6dF \cite{Ross:2014qpa} and SDSS DR16 \cite{eBOSS:2020yzd}. The last also includes redshift distortion (RSD) measurement of $f\sigma_8$.
\end{itemize}
The main constraining power on $P_{\rm m}(k,z)$ comes from weak lensing and galaxy clustering as measured in LSS, in particular lensing convergence and redshift space distortions (RSD), and CMB lensing. Under the Limber approximation,  and the Newtonian approximation for the metric potentials,  
 the lensing convergence spectrum for two tomographic redshift bins $i,j$ is
\begin{equation}\label{eq:clkk}
    C^{\kappa_i\kappa_j}_\ell=\left(\frac{3}{2}\Omega_mH_0^2\right)^2\int \frac{d\chi}{a^2(\chi)}g_L^i(\chi)g_L^{j}(\chi)P_{\rm m}\left(\frac{\ell + 1/2}{\chi},z(\chi)\right),
\end{equation}
with the lensing kernel
\begin{equation}\label{eq:lensing kernel}
    g_L^i(\chi)=\int_\chi^{\chi_H}d\chi' n_i(\chi')\frac{\chi-\chi'}{\chi'},
\end{equation}
where $\chi_H$ is the conformal radius of the horizon today and $n_i$ is the redshift distribution of sources in the $i-th$ redshift bin. The LSS likelihood handles $n_i$ internally and takes $P_{\rm m}(k)$ as input, (with $k=(\ell+1/2)/\chi$). CMB lensing has a single source at the last scattering surface thus $n_{\rm CMB}(\chi)=\delta(\chi-\chi_*)$ with $\chi_*$ being the conformal distance to the last scattering surface. The CMBL data takes the CMB lensing convergence spectrum computed from \eqref{eq:clkk} as input. Fig.\ref{fig:lensing kernel} plots the lensing kernel \eqref{eq:lensing kernel} for LSS and CMBL, showing that LSS and CMBL constrain $P_{\rm m}$ at different redshifts. To further constrain the reconstruction one also needs to reduce the uncertainty in the linear growth rate $D_g$. To this end, we use the SNIa and BAO data to constrain the background evolution.

To sample the cosmology and reconstruct the matter power spectrum, we perform a  Monte Carlo Markov chain (MCMC) analysis using the cosmology sampler \texttt{Cobaya} \cite{Torrado:2020dgo,2019ascl.soft10019T} over the dataset LSS+CMBL+SN+BAO. All likelihood codes are available with the public version of \texttt{Cobaya} except for the DES Y3 3x2pt LSS likelihood, which we ported from \texttt{CosmoSIS} \cite{Zuntz:2014csq} into \texttt{Cobaya} as new modules \footnote{Publicly available at \url{https://github.com/JiangJQ2000/cosmosis2cobaya}}. We use the Gelman-Rubin test \cite{Gelman:1992zz} $R-1<0.02$ as our convergence criterium. Since we focus on LSS, and use only the lensing of CMB, the primordial curvature spectrum tilt $n_s$, the baryon density parameter $\omega_b=\Omega_b h^2$ and the effective optical depth $\tau$ are poorly constrained. We therefore fix the spectrum tilt and optical depth to the fiducial values $n_s=0.966$, $\tau=0.055$ and apply a BBN prior $\omega_b=0.02218\pm0.00055$ \cite{Schoneberg:2024ifp} on the baryon density parameter. Following Planck we treat neutrinos as two massless and one massive with mass 0.06 eV, reproducing $N_{\rm eff}=3.044$ \cite{Bennett:2020zkv,Froustey:2020mcq,Akita:2020szl}.

In the zeroth order expansion, the coefficient $A_0(k)$ has an overall scaling freedom degenerate with the primordial scalar perturbation amplitude $A_s$. We thus fix the first node $\ln A_0(k_1)=0$ (assuming no correction to the spectrum deeply in the linear regime) and the last node $\ln A_0(k_6)=-\sum_{i=1}^{5}\ln A_0(k_i)$ (removing the overall scaling by forcing a vanishing average). For $A_1(k)$ we fix the first two nodes to a vanishing value $\ln A_2(k_{1,2})=-20$ because non linear correction is irrelevant in the linear regime. In conclusion, the model parameters sampled in the MCMC are $\{H_0, \omega_b, \omega_{\rm cdm}, A_s, \ln A_0(k_{2,3,4,5}), \ln A_1(k_{3,4,5,6}) \}$, with the prior choice reported in Table.\ref{tab:prior}. In addition to the model parameters, we also vary all nuisance parameters of each likelihood with their recommended priors. Note that for LSS we use the nuisance setup from Ref.~\cite{Kilo-DegreeSurvey:2023gfr}. In particular we use the NLA-z model instead of TATT for linear-alignment template as suggested by Ref.~\cite{Kilo-DegreeSurvey:2023gfr}.

\begin{table}
    \begin{tabular}{cc}
    \hline
    \hline
    \multicolumn{2}{c}{Priors}\\
    \hline
    $H_0$&$\mathcal{U}[64,82]$\\
    $\omega_{\rm b}$&$\mathcal{N}(0.02218, 0.00055)$\\
    $\omega_{\rm cdm}$&$\mathcal{U}[0.051,0.255]$\\
    $10^9A_s$&$\mathcal{U}[0.5,5]$\\
    $\ln A_0(k_{2,3,4,5})$&$\mathcal{U}[-5,5]$\\
    $\ln A_1(k_{3,4,5,6})$&$\mathcal{U}[-20,20]$\\
    \hline
    \end{tabular}
\caption{Priors used for the MCMC parameters.}
\label{tab:prior}
\end{table}

\section{Results}\label{sec:result}

\begin{figure}
    \centering
    \includegraphics[width=0.49\linewidth]{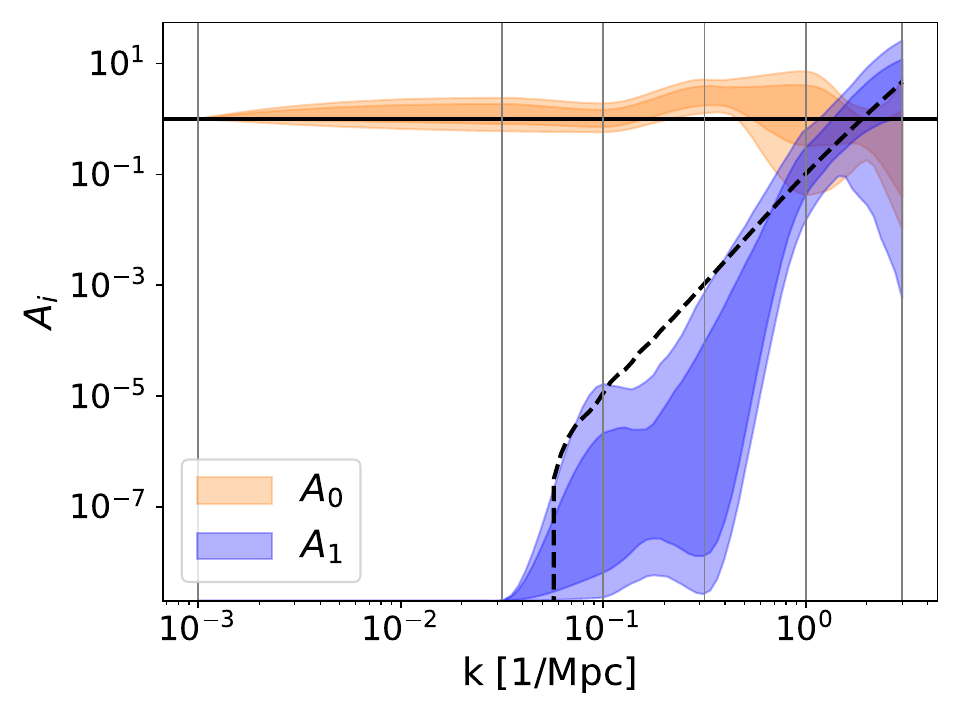}
    \includegraphics[width=0.49\linewidth]{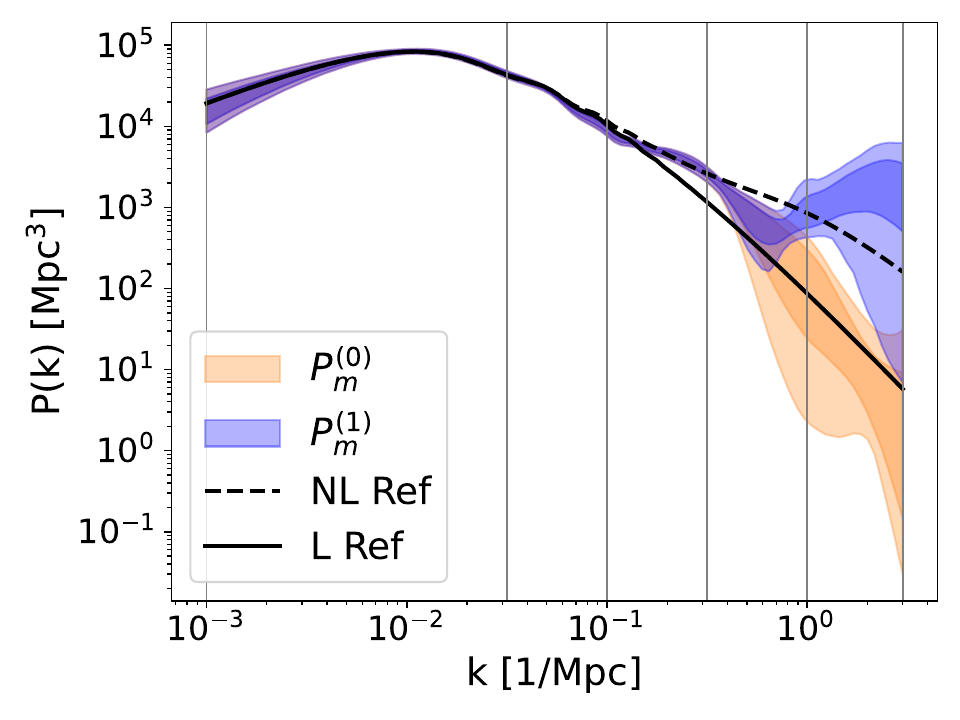}
    \caption{{\bf Reconstruction of the Power Spectrum} - Shaded contours depict 68\% and 95\% confidence intervals of the function values at each $k$. Vertical gray lines mark the position of the interpolation nodes. \textit{Left panel:} the reconstructed $A_0(k)$ and $A_1(k)$.  \textit{Right panel:} the reconstructed matter power spectrum. The yellow contours show the leading order part of the reconstructed spectrum, i.e. $A_0(k)P_{\rm L}(k)$ while the blue contours show the complete reconstruction. Black solid and dashed lines refer to $P^{\rm ref}_{\rm L}(k)$ and $P^{\rm ref}_{\rm NL}(k)$,  respectively.}
    \label{fig:a0a1}
\end{figure}

\begin{figure}
    \centering
    \includegraphics[width=\linewidth]{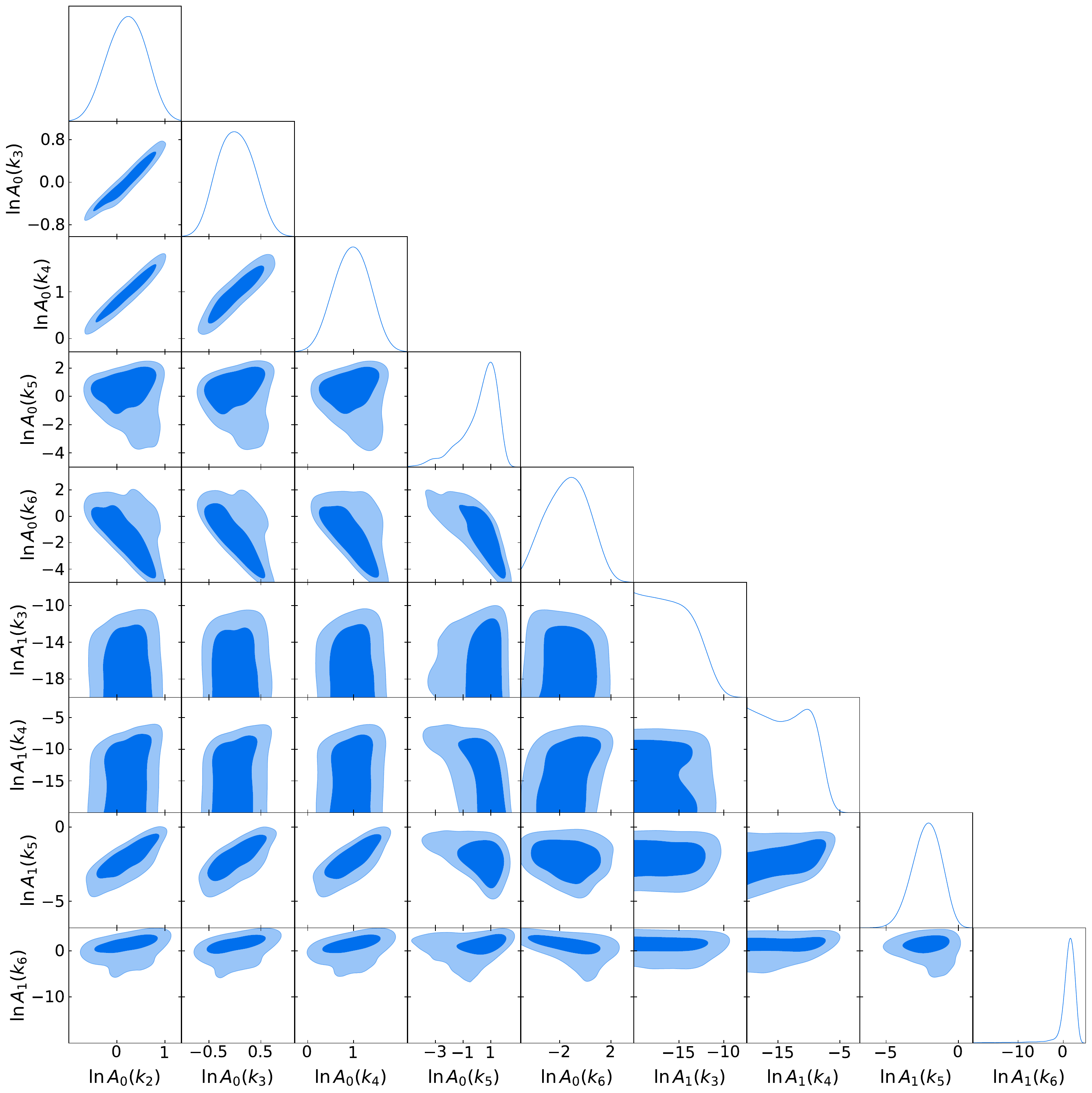}
    \caption{68\% and 95\% marginalized posteriors of the interpolation nodes for $A_0(k)$ and $A_1(k)$.}
    \label{fig:trig_node}
\end{figure}

The results of the reconstruction of the free functions $A_i(k)$, as well as the resulting power spectra up to first order are plotted in Fig.\ref{fig:a0a1}, while the posteriors of the nodes for the $A_i(k)$ are reported in Fig.\ref{fig:trig_node}. In the same figure, we also plot the reference linear power spectra, $P_{\rm L}^{\rm ref}$, and non-linear, $P_{\rm NL}^{\rm ref}$ respectively, corresponding to the $\Lambda$CDM bestfit point using the same datasets as the reconstruction, with non-linear correction by~\texttt{HMCode2020} \cite{Mead:2020vgs}. We can compare these with the reconstruction at zero-th and $n$-th order, respectively; a good agreement signals that the proposed method, at $n$-th order is able to reconstruct the full $P_m(k,z)$ within the range of scales relevant for the data under consideration. We also compare each reconstructed $A_i(k)$, with their values in the reference model, i.e. $A^{\rm ref}_0(k)=1$ and $A^{\rm ref}_1(k)\equiv(P^{\rm ref}_{\rm NL}(k)-P^{\rm ref}_{\rm L}(k))/[P^{\rm ref}_{\rm L}(k)]^2\,{\rm Mpc}^{-3}$ \footnote{Precisely, $P_{\rm NL}=\sum_{n\ge1}P^{(n)}_{\rm m}$ includes all $A_n$ with $n\ge1$. However, as commented in the main text, cutting off at $n=1$ is enough for the datasets used thus $(P^{\rm ref}_{\rm NL}(k)-P^{\rm ref}_{\rm L}(k))/[P^{\rm ref}_{\rm L}(k)]^2\,{\rm Mpc}^{-3}$ is a good approximation of $A^{\rm ref}_1$ at current precision.}. Since the non-linear correction is not computed for $k<0.1 \,{\rm h}\, \rm{Mpc}^{-1}$, the  line for $A^{\rm ref}_1(k)$ drops vertically at linear scales, as can be seen in the left panel of Fig.\ref{fig:a0a1}. 
$A_1(k)$ becomes relevant between the fourth and fifth-node, corresponding to $0.3<k<1\,{\rm Mpc}^{-1}$; in that range, one can observe from Fig.~\ref{fig:a0a1} that while in the left panel the $A_1(k)$ (blue) stripe is generally below the reference black dashed line, the contour of $A_0$ has a bump above the solid black line, causing the $P^{(0)}_{\rm m}$ stripe to overlap  with the full spectrum (dashed line) in the right panel. This implies that up to the fourth node ($k\lesssim0.3 \ \rm{Mpc}^{-1}$), the reconstruction cannot distinguish between the $A_0$ and $A_1$ terms of Eq.\eqref{eq:pk_expansion2} and the linear term, $A_0(k)$, is incorrectly distorted to fit the full non-linear spectrum.

First, let us remark that since in Eq.\eqref{eq:pk_expansion2}, we have a theoretical modeling of $P_{\rm L}(k,z)$, i.e. $\Lambda$CDM, the constraints on $A_1(k)$ and $P_{\rm m}^{(0)}$ are generally good, as expected. It can be noticed that the data provide also good constraints on $P_{\rm m}(k)$ between the second and last nodes, corresponding to the interval $k\in (0.03, 3) \ \rm{Mpc}^{-1}$. Especially, the reconstruction recovers the full $P_{\rm m}(k)$ without assuming any non-linear model. In particular, the last two nodes of $A_1$ in Fig.~\ref{fig:trig_node} display clear peaks away from their prior boundaries despite the fact that the reconstruction does not assume any  theoretical modeling of the non-linear part and uses wide priors on $\ln A_1$ crossing 17 orders of magnitude. This implies that our method based on Eq.\eqref{eq:pk_expansion2} successfully reconstructs both the linear and non-linear spectrum directly from data. This can also be seen in Fig.~\ref{fig:a0a1}, where in the left panel the $A_1$ (blue) contour correctly reproduces the black dashed line obtained from the Halo model, while in the right panel, the $P^{(0}_{\rm m}$ (yellow) and $P^{(1)}_{\rm m}$ (blue) strips disjoin to fit the linear and non-linear spectra respectively. As $A_1(k)$ already requires $k\gtrsim1 \ \rm{Mpc}^{-1}$ to be well constrained, it is to be expected that higher order terms in Eq.\eqref{eq:pk_expansion} and Eq.\eqref{eq:pk_expansion2} cannot be constrained by current data. We confirm this expectation by further opening up $A_2(k)$ in Appendix-\ref{apdx:more results}.

Let us conclude this Section by looking at the clustering parameter $S_8$. The latter is commonly  defined from the  linear matter power spectrum, which in our case corresponds to $P^{(0)}_{\rm m}$ from Eq.\eqref{eq:pk_expansion2}. We therefore compute $S_8$ using $P^{(0)}_{\rm m}$ from the reconstruction finding  the posterior $S_8=0.88\pm0.029$, which is significantly larger than the $S_8=0.772^{+0.018}_{-0.017}$ reported by~\cite{DES:2021bvc, DES:2021vln} using the DES Y3 data. It is also larger than the $S_8=0.831\pm0.029$ from~\cite{Carron:2022eyg} based on Planck PR4 lensing and BAO. This is expected because, as noted in the previous paragraphs, at the scales relevant for $S_8$, i.e. $k\sim 0.125h \ \rm{Mpc}^{-1}$, the reconstruction confuses $A_0$ with the non-linear correction, see also Fig.~\ref{fig:a0a1}. As a result, the $S_8$ obtained is in fact computed from the full matter spectrum rather than the linear one and is correspondingly  larger. Within the dataset that we use, RSD constrains $f\sigma_8$ directly, so it could in principle drag the $S_8$ posterior to lower values, however its constraining power is not strong enough to compete with the other datasets in consideration. Posteriors of the cosmological parameters are shown in Appendix-\ref{apdx:more results}.

\section{Discussion}
In this work, we proposed a new model-independent reconstruction method of the matter power spectrum based on its time dependence, Eq.\eqref{eq:pk_expansion} and \eqref{eq:pk_expansion2}, and observations from different redshifts. Using DES Y3 and Planck PR4 CMB lensing, we found that the method can reconstruct the shape of the full matter power spectrum from current observations and further simultaneously reconstruct its linear and non-linear parts for $k\gtrsim 1\ \rm{Mpc}^{-1}$ without assuming any model for the non-linear scales. 

The results presented in this paper are only a proof of concept. Within the $\Lambda$CDM paradigm, the reconstruction can serve as a cross check of the modeling of the non-linear regime, as shown in Fig.~\ref{fig:a0a1}. From another point of view, deviations from the $\Lambda$CDM expectation in the reconstruction might also be an indication of beyond $\Lambda$CDM physics. The natural next step is to study whether the proposed reconstruction can shed light on assessing the impact of baryonic feedback in the non-linear regime with more data, e.g. cross correlated galaxy and CMB lensing~\cite{DES:2022urg}, and more precise future data from e.g. stage-IV CMB~\cite{CMB-S4:2016ple} and LSS~\cite{Euclid:2024yrr,LSST:2008ijt}. We have observed that current data only provides constraints up to $n=1$ in Eq.\eqref{eq:pk_expansion2}. It will be interesting to assess whether Stage IV data can constrain $n\ge2$ with mock data.

Interestingly, for beyond $\Lambda$CDM cosmology with dynamical dark energy and/or modified gravity where non-linear evolution is either not fully understood or simply too time-consuming to resolve model by model when scanning the theory space, the reconstruction Eq.\eqref{eq:pk_expansion2} can provide an alternative way of dealing with the non-linear corrections while using only knowledge about the linear evolution. Like for the $\Lambda$CDM case, the reconstruction can further serve as a consistency check for existing non-linear modeling in the said model. For example, Fig.~\ref{fig:a0a1}, shows that the reconstruction can provide information approximately on  the scale at which non-linear growth (effects that cannot be fitted with $A_0(k)$) becomes important to observations, which will be interesting to study in the dark energy / modified gravity models. Furthermore, it has recently been observed that modified gravity effect can be degenerate with baryonic effects \cite{Euclid:2024xfd}. The ability to reconstruct linear and non-linear matter power spectrum by its time dependence might provide a potential way to break this degeneracy.

\begin{acknowledgments}
The authors thank Daan Meerburg for insightful comments. GY and AS acknowledge support from the NWO and the Dutch Ministry of Education, Culture and Science (OCW) (through NWO VIDI Grant No. 2019/ENW/00678104 and ENW-XL Grant OCENW.XL21.XL21.025 DUSC) and from the European Research Council under the H2020 ERC Consolidator Grant “Gravitational Physics from the Universe Large scales Evolution” (Grant No. 101126217 — GraviPULSE).
\end{acknowledgments}

\appendix

\section{More MCMC results}\label{apdx:more results}
\begin{figure}
    \centering
    \includegraphics[width=\linewidth]{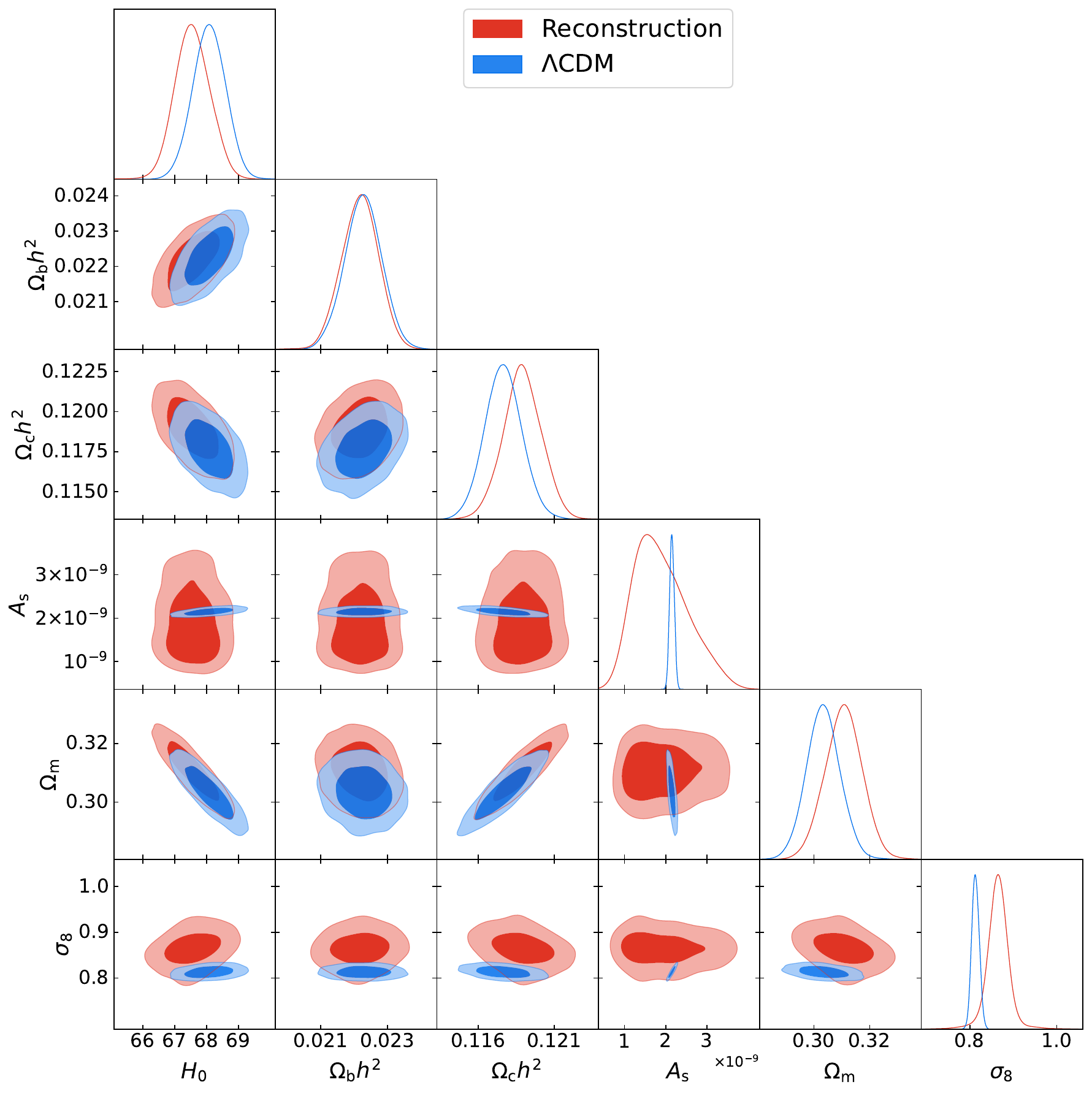}
    \caption{68\% and 95\% marginalized posteriors of the cosmological parameters for the $\Lambda$CDM reference model and the reconstructed model against the same datasets as described in the main text.}
    \label{fig:trig_cosmo}
\end{figure}

Fig.~\ref{fig:trig_cosmo} plots the 68\% and 95\% posteriors of the cosmological parameters for the reconstruction studied in the main text. The posterior constraint on the primordial curvature perturbation amplitude, $A_s$, is significantly enlarged in the reconstruction compared to the $\Lambda$CDM case. This is not surprising because the $A_0$ coefficient in Eq.\eqref{eq:pk_expansion2} is degenerate with $A_s$ for the datasets we considered. The reason for the increased $\sigma_8$ is explained in detail in the main text, i.e. the reconstruction incorrectly uses the linear spectrum, $P^{(0)}_{\rm m}$, to fit the full non-linear $P_{\rm m}$ around the scales relevant to $\sigma_8$.

\begin{figure}
    \centering
    \includegraphics[width=\linewidth]{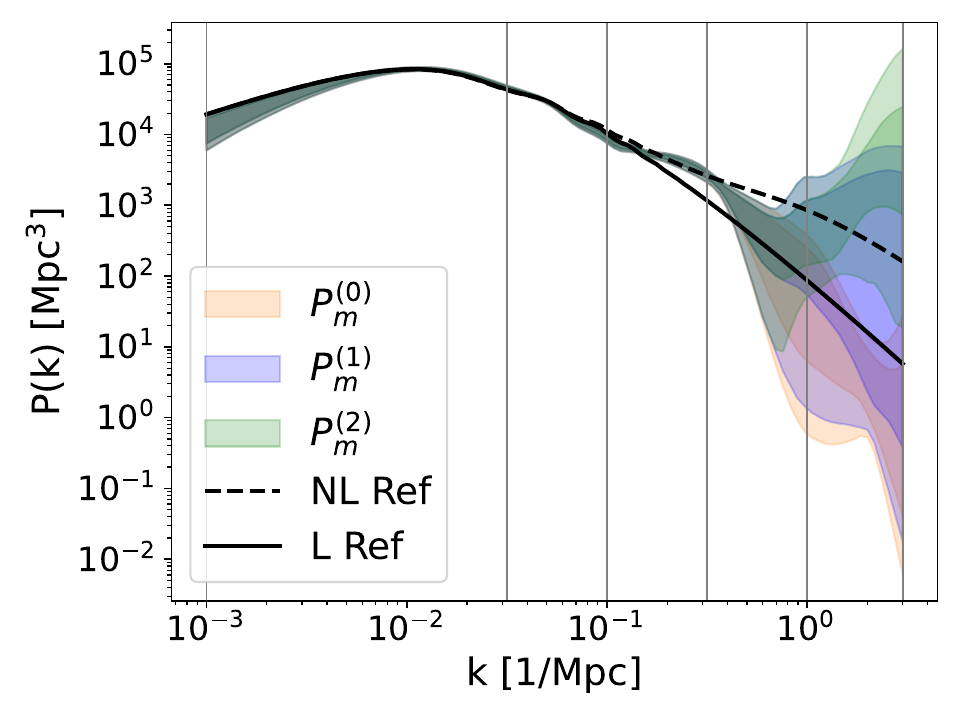}
    \caption{Reconstructed $P_{\rm m}(k)$ expanding up to $P_{\rm m}^{(2)}$.}
    \label{fig:a0a1a2}
\end{figure}

\begin{figure}
    \centering
    \includegraphics[width=\linewidth]{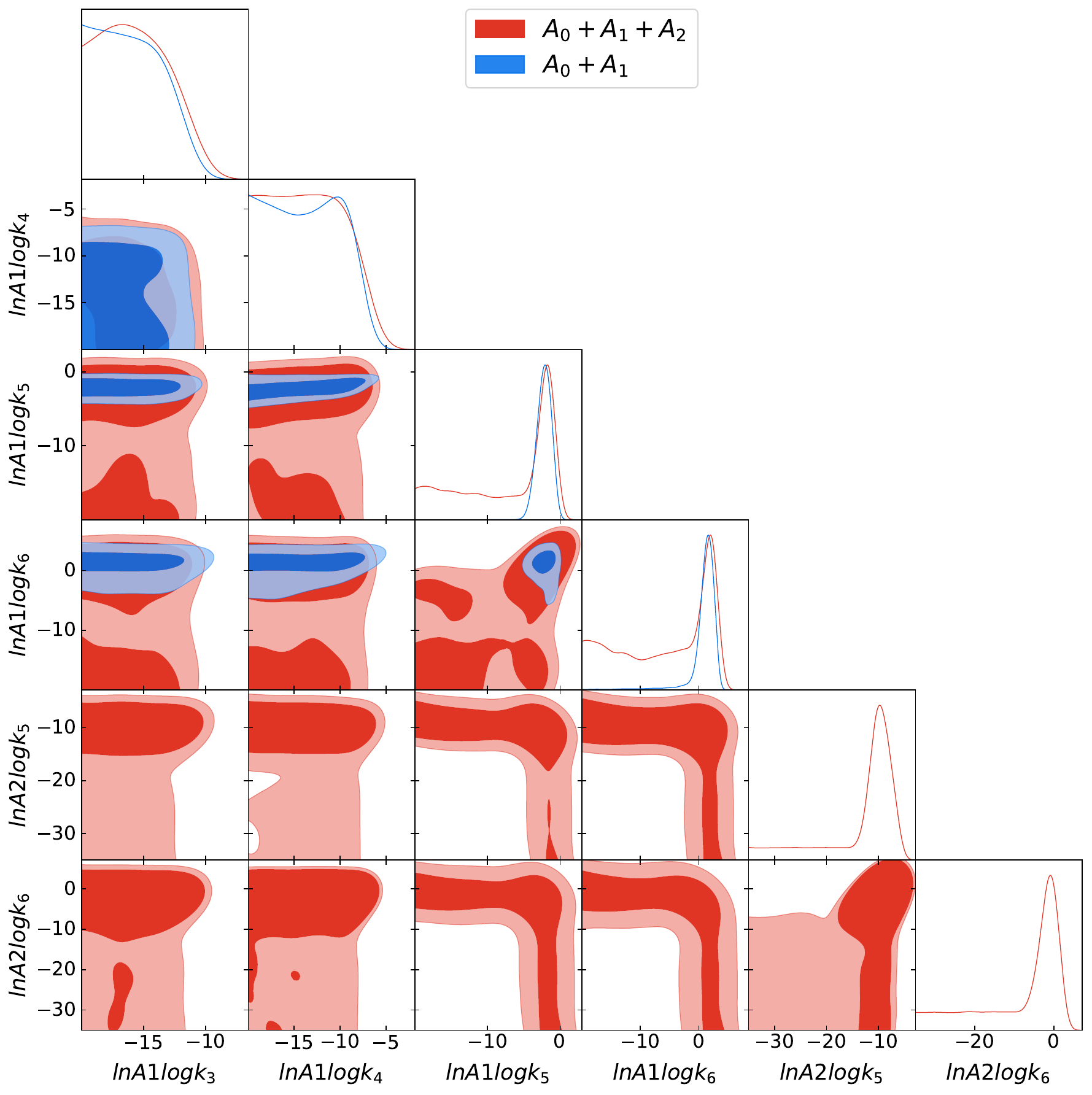}
    \caption{68\% and 95\% marginalized posterior of the interpolation nodes for $A_1(k)$ and $A_2(k)$.}
    \label{fig:trig_a3}
\end{figure}

In the main text, we cut off Eq.\eqref{eq:pk_expansion2} at $n=1$. Here we go up to $n=1$, i.e. include also  $A_2(k)$, with a uniform prior $-35<\ln A_2(k_{5,6})<20$. From the result in the main text for the $n=1$ case, we know that $A_1$ is only constrained for $k\gtrsim 1 \ \rm{Mpc}^{-1}$ thus we fix $\ln A_2(k_{1,2,3,4})=-35$ without loss of generality. The node posterior results are plotted in Fig.~\ref{fig:trig_a3}.
Note that the MCMC analysis had great difficulty converging in reasonable time when $A_2$ is also free, thus we plotted here the result with $R-1\approx0.3$. It is clear from the figure, as well as the difficulty in convergence, that $A_1$ and $A_2$ nodes are highly degenerate. Furthermore, Fig.~\ref{fig:a0a1a2} plots the reconstructed $P_{\rm m}(k)$, expanding up to $P_{\rm m}^{(2)}$, in the same style as the right panel of Fig.~\ref{fig:a0a1}. One can observe that the reconstructed shape (green) is essentially consistent with the result with only $A_0$ and $A_1$, but suffers serious degeneracy as depicted in Fig.~\ref{fig:trig_a3}. We therefore conclude that the current data cannot distinguish $A_1$ from $A_2$ even at the smallest scales.

\bibliography{free_pk}

\end{document}